\documentclass{optica-article}

\journal{opticajournal} 

\articletype{Research Article}

\usepackage{lineno}
\usepackage[normalem]{ulem}

\begin{document}

\title{Pulse Breathing Dynamics in a Mode-Locked Laser measured via SHG autocorrelation}

\author{S. Kannan, S. Padmanaban, X. T. Yan, Y. S. Athreya,\\ K. G. H. Baldwin, S. S. Hodgman, and A. G. Truscott\authormark{*}}

\address{Department of Quantum Science and Technology, Research School of Physics, The Australian National University, Canberra, ACT 2601, Australia.}

\email{\authormark{*}andrew.truscott@anu.edu.au} 


\begin{abstract*} 
Pulse-to-pulse fluctuations in mode-locked lasers fundamentally limit applications from optical frequency combs to supercontinuum generation. While timing jitter has been extensively characterized, pulse amplitude and width fluctuations remain less accessible experimentally. We present a statistical autocorrelation method that demonstrates pulse breathing dynamics through Fano factor analysis of second-harmonic generation autocorrelation.
\textcolor{black}{This reveals a characteristic W-shape in the enhanced Fano profile, a signature of pulse shape dynamics that is invisible to time-averaged fluctuations. Applying this method to two commercially available passively mode-locked oscillators operating at 1030 nm and 1045 nm, with different performance specifications, we measure pulse width fluctuations of 3.2(1)\,fs and 2.86(2)\,fs respectively. The two independent instruments serve as a cross-validation of the technique across different laser platforms.}
This diagnostic capability opens the door to identifying and suppressing specific breathing mechanisms, paving the way for the design of ultra-stable oscillators required for precision frequency metrology.
\end{abstract*}

\section{Introduction}
Passively mode-locked lasers capable of generating femtosecond pulse trains have become the corner stone of modern ultrafast photonics, with applications spanning optical frequency combs\,\cite{fortier201920,udem2002optical}, precision spectroscopy\,\cite{diddams2007molecular}, high-capacity optical communications\,\cite{pfeifle2014coherent}, femtochemistry and femtobiology\,\cite{wang2011progress}, and attosecond science\,\cite{RevModPhys.81.163}. Recent advances in solid-state gain media, particularly Yb-doped crystals, have further enabled the development of compact, highly efficient sources operating in the near infrared region\,\cite{Saraceno:14,Pirzio:16}. These systems increasingly serve as sources for nonlinear frequency conversion, such as supercontinuum generation in photonic crystal fibers\,\cite{RevModPhys.78.1135} and high-harmonic generation of extreme-UV sources\,\cite{UV-femto}. The efficacy of these applications depends critically on the stability of pulse evolution\,\cite{PhysRevLett.90.113904,corwin2003fundamental}. Specifically, in supercontinuum generation, minute fluctuations in the pump pulse width can translate into macroscopic jitter in the soliton fission length, leading to severe decoherence and intensity noise at the spectral edges\,\cite{PhysRevLett.90.113904}. Similarly, in optical frequency combs, pulse shape fluctuations couple directly to phase noise via the Kerr effect (amplitude-to-phase conversion), ultimately limiting the achievable carrier-envelope phase (CEP) stability and comb linewidth. While often assumed to be stationary, mode-locked pulses can exhibit complex transient dynamics, motivating the need for accessible techniques to characterize these pulse-to-pulse width fluctuations. 

Standard pulse characterization techniques such as frequency-resolved optical gating (FROG)\,\cite{Trebino:93,DeLong:94,trebino} and spectral phase interferometry (SPIDER)\,\cite{Iaconis:98} can be used to retrieve the time averaged electric field and spectral phase. However, these integrating measurements average out rapid dynamics, obscuring pulse-to-pulse fluctuations. While timing jitter (stochastic variation in pulse arrival time) has been characterized to sub-femtosecond precision using balanced optical cross-correlation\,\cite{Kim:07,Schibli:03} and is well-understood theoretically\,\cite{206583,paschotta2004noise,paschotta2004noise2}, pulse shape fluctuations remain poorly quantified. Although single-shot spectroscopic methods exist\,\cite{Kane:93,Salin:87,goda2013dispersive}, they require pulse energies in the microjoule range and complex setups. Recently, Barbero et al.\,\cite{Barbero_2025} demonstrated single-shot characterization of shot-to-shot pulse-shape fluctuations using birefringent wedges and an imaging spectrometer, enabling individual pulse retrieval from unstable sources. While such approaches provide complete temporal characterization of each pulse, they require high pulse energies and expensive imaging spectrometers, rendering them unsuitable for the routine characterization of nanojoule level oscillators. For stable oscillators operating at nanojoule pulse energies, direct statistical methods offer a more practical alternative. This diagnostic gap is particularly critical for soliton mode-locked lasers, where the soliton area theorem dictates a fundamental coupling between pulse energy and duration\,\cite{AGRAWAL2013129}. Consequently, even minor amplitude noise can drive correlated breathing in the pulse-width dynamics that remain invisible to conventional time-averaging diagnostics.  Resolving these fluctuations provides a direct probe into the internal cavity physics, revealing how quantum noise and gain dynamics perturb the soliton attractor. This allows for the characterization of the laser’s specific noise transfer function, distinguishing between simple additive noise and coupled nonlinear breathing modes.

In this work, we demonstrate that statistical analysis of standard second-harmonic generation (SHG) autocorrelation signals provides a direct probe of pulse breathing dynamics. By acquiring thousands of single shot SHG intensity traces at each discrete delay position and computing the Fano factor (variance-to-mean ratio), we quantify a measure of pulse width fluctuations. This technique exploits the quadratic intensity dependence of the SHG process\,\cite{boyd2008nonlinear}, where the signal variance becomes highly sensitive to pulse width variations especially away from the pulse center. \textcolor{black}{We apply this method to two mode-locked oscillators operating at 1030 nm and 1045 nm respectively.} Finally, we utilize these retrieved fluctuation parameters to predict the coherence degradation in supercontinuum generation\,\cite{Genier:19}.

The paper is organized as follows. Section 2 develops the theoretical framework, deriving the variance decomposition formula and defining the Fano enhancement ratio. Section 3 describes the experimental apparatus and measurement protocol. Section 4 presents results including FROG characterization of the average pulse and statistical autocorrelation revealing the characteristic shape of the enhanced Fano profile. Section 5 discusses physical interpretation and implications for applications. Section 6 concludes.

\section{Theoretical Framework}
In a pulsed laser operating in the soliton mode-locking regime, pulse parameters are coupled through the interplay of dispersion, nonlinearity, gain, and loss\,\cite{902165}. 
The resulting temporal intensity profile $I(t)$ is well-described by a hyperbolic secant squared function:
\begin{equation}
I(t) = P_0 \text{sech}^2\left(\frac{t}{\tau_0}\right),
\label{eq:pulse_shape}
\end{equation}
where $P_0$ is the peak power and $\tau_0$ is the pulse duration parameter. The well known soliton area theorem relates energy $(E\propto P_0 \tau_0)$ to the cavity parameters and states that the product $E\times\tau_0$ is an invariant of the system\,\cite{10.1063/1.321997}. Under first-order perturbations to the cavity parameters, the pulse must instantaneously adjust its shape to maintain the soliton condition. This predicts a strong anti-correlation (Pearson coefficient $\rho\approx-1$) between energy fluctuation and pulse-width fluctuations.

\begin{figure}[t]
    \centering
    \begin{minipage}{\textwidth}
        \centering
        \includegraphics[width=\linewidth,height=4cm]{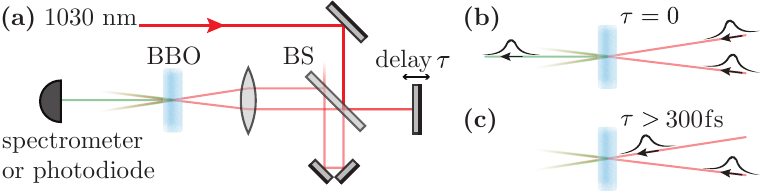}
    \end{minipage}
    \caption{ (a) Schematic of the non-collinear intensity autocorrelator. The input pulse is split into two replicas by a beam splitter (BS); one arm traverses a variable delay line ($\tau$) before both are focused into a BBO crystal. The background-free sum-frequency generation (SFG-green line) energy is measured as a function of delay to yield the autocorrelation trace. (b) Illustration of the interaction at zero delay ($\tau=0$): the pulses overlap temporally and spatially, generating the SFG signal (central beam). (c) At large delays (e.g., $\tau>300\,$fs), the pulses are temporally separated, and no SFG signal is produced. BS$=50:50\,$non-polarizing beamsplitter, BBO$=\beta-$barium borate.}\label{fig1}
\end{figure}

\textcolor{black}{To precisely capture these coupled dynamics, we employ a statistical analysis of the SHG autocorrelation (see Fig.\,\ref{fig1}). For two delayed pulse replicas separated by delay $\tau$, the SHG signal is proportional to $I(t)\cdot I(t-\tau)$. Since the SHG pulse duration ($\sim 200\,$fs) is much smaller than the APD detector response ($\sim 1\,$ns at 1\,GHz bandwidth), each electrical detector pulse represents the time-integrated optical energy of the corresponding SHG pulse.}

\textcolor{black}{To extract the breathing parameter, we compute the total area under the digitized electrical pulse for each shot, calculated as the sum of the sampled voltage values weighted by the ADC sampling period:
\begin{equation}\label{dissum}
    S(\tau)=\Delta T_s\sum_nV_n(\tau),
\end{equation}
where $\Delta T_s=160\,$ps is the ADC sampling period of the oscilloscope and $V_n$ is the voltage value of the $n^\text{th}$ sample. The sum runs over all samples within a fixed time window enclosing the detector pulse. This integrated signal is directly proportional to the SHG autocorrelation:
\begin{equation}
    S(\tau)\propto \int I(t)\cdot I(t-\tau) dt.
\end{equation}
}
Since the autocorrelation of a sech$^2$ pulse is also a sech$^2$ function (with a width expansion factor), we model the time-averaged intensity of the autocorrelation signal $S_{ac}(\tau)$ as:
\begin{equation}
    S_{ac}(\tau, w)\simeq S_0\text{sech}^2(\tau/w),
\end{equation}
where $S_0$ represents the peak SHG intensity, $w$ is the width parameter, and $\tau$ is the delay. The exact analytical autocorrelation of a hyperbolic secant pulse is described by a transcendental function involving $\coth(\tau)$ terms, although it is widely approximated by a $\text{sech}^2$ profile due to the negligible deviation between the two forms ($<1\%$ root-mean-square error). Under this approximation, the full-width at half-maximum (FWHM) of the autocorrelation trace is related to the pulse duration $\tau_0$ by the deconvolution factor $\tau_{ac}\approx 1.543\tau_0$. Hence, we treat the autocorrelation signal as a sech$^2$ function to derive closed-form expressions for the sensitivity derivatives.

\begin{figure}[t]
    \centering
    \begin{minipage}{0.48\textwidth}
        \centering
        \includegraphics[width=\linewidth,height=5.5cm]{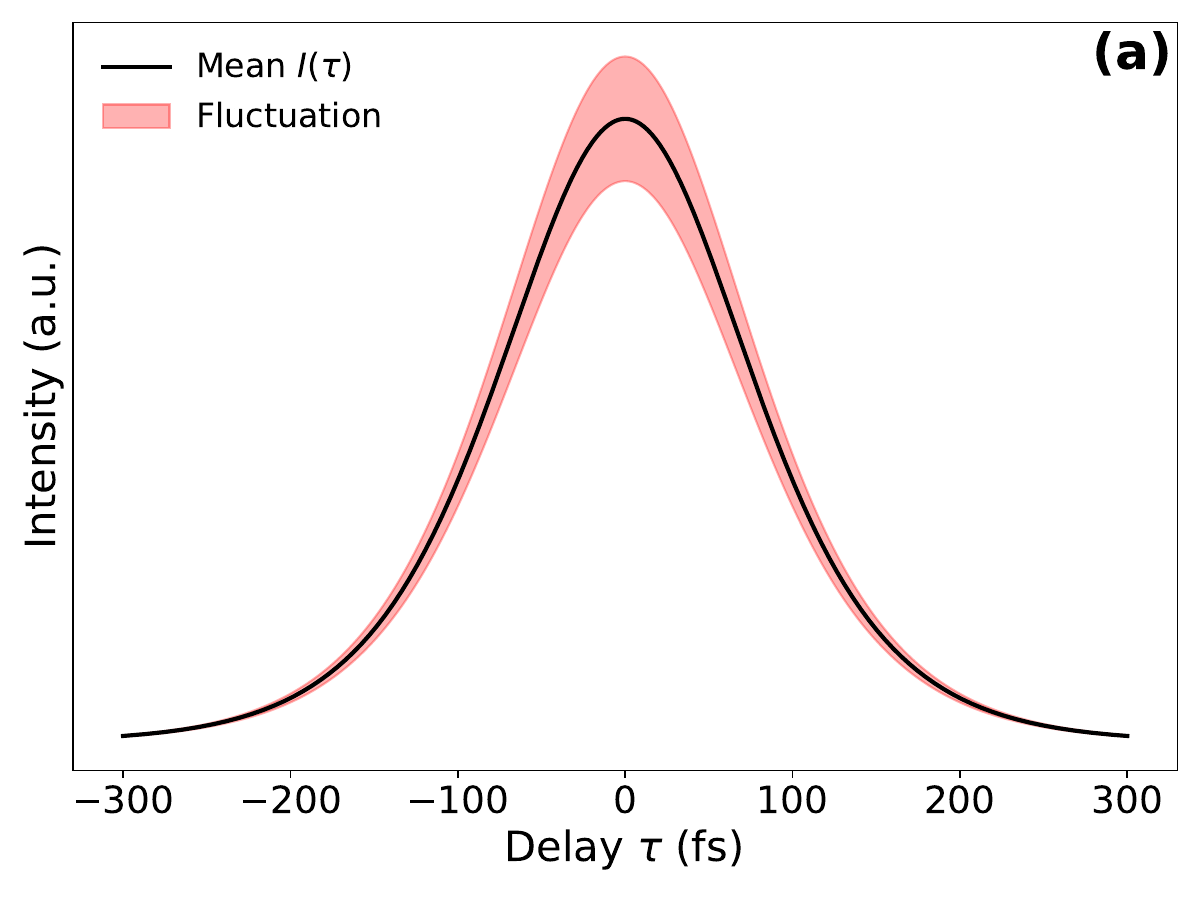}
    \end{minipage}
    \begin{minipage}{0.48\textwidth}
        \centering
        \includegraphics[width=\linewidth,height=5.5cm]{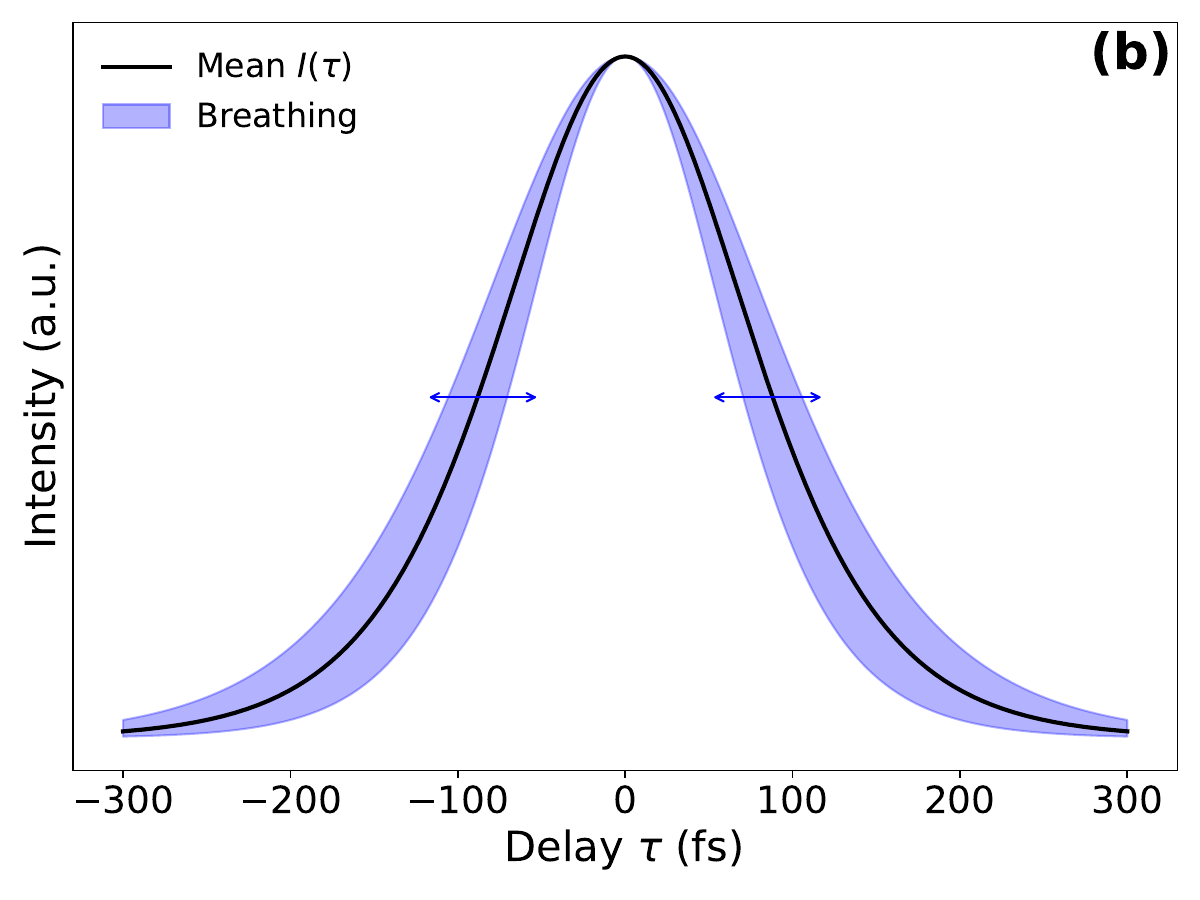}
    \end{minipage}
    \caption{Schematic representation of the noise contributions derived in Eq.\,\eqref{eq:variance_prop}. (a) Amplitude noise causes intensity fluctuations that scale with the pulse envelope, maximizing at zero delay ($\tau=0$). (b) Pulse breathing (width fluctuations) changes the pulse duration. Crucially, this creates negligible intensity change at the pulse peak ($\tau=0$) and far wings, but maximum change at the pulse shoulders.}\label{amp_breath_noise}
\end{figure}

Assuming small perturbations, the variance of the intensity fluctuations $\langle \Delta S(\tau)^2 \rangle$ at any delay position can be expanded using first-order error propagation \textcolor{black}{by incorporating fluctuations in the amplitude baseline, pulse width (breathing), and mechanical stage jitter:}
\begin{equation}\label{eq:variance_prop}
    \langle \Delta S(\tau)^2 \rangle \approx \langle S(\tau)\rangle^2 \Big[\mathcal{R}^2+k(\tau)^2\left(\frac{\sigma_w}{w}\right)^2+j(\tau)^2\left(\frac{\sigma_{\tau}}{w}\right)^2+2\rho \mathcal{R} k(\tau)\left(\frac{\sigma_w}{w}\right)\Big],
\end{equation}
\textcolor{black}{where $\mathcal{R}$ is the relative intensity noise (RIN) of the zero-delay signal, $\sigma_w$ is the standard deviation of shot-to-shot width fluctuations, and $\sigma_\tau$ is the RMS stage positioning noise assumed uncorrelated with the laser parameters (see Fig.\,\ref{amp_breath_noise})}. $k$ and $j$ are the dimensionless sensitivity factors. The breathing sensitivity factor $k(\tau)$ is defined as
\begin{equation}
    k(\tau)=\frac{1}{\langle S(\tau)\rangle}\frac{1}{w}\frac{\partial S}{\partial w}.
\end{equation}
The stage noise term is distinguished from pulse breathing by an analogous sensitivity function. A shot-to-shot stage position jitter $\delta\tau$ produces an intensity fluctuation $\delta S=(\partial S/\partial\tau)\delta \tau$. Normalizing by the local mean intensity and factoring out $(\sigma_{\tau}/w)$ in the same manner as $k(\tau)$, we define $j(\tau)$: 
\begin{equation}
j(\tau) = \Big(\frac{w}{\langle S(\tau)\rangle}\Big)\frac{\partial S}{\partial\tau}.
\end{equation}
\textcolor{black}{We note that at zero delay the breathing sensitivity $k(0)=0$ and width fluctuations contribute no variance beyond the baseline $\mathcal{R}^2$.}

To get a sensitive qualitative diagnostic of pulse breathing, we calculate the Fano factor. It is defined as the ratio of variance-to-mean $F(\tau) = \langle \Delta S(\tau)^2 \rangle / \langle S(\tau) \rangle$. \textcolor{black}{Using this Fano factor, we define the dimensionless enhancement factor ($\varepsilon(\tau)$) by normalizing the measured variance by the amplitude noise baseline $\mathcal{R}^2\langle S(\tau)\rangle^2$:
\begin{equation}\label{enhan}
    \varepsilon(\tau) = \frac{\langle \Delta S(\tau)^2 \rangle}{\mathcal{R}^2\langle S(\tau) \rangle^2}\approx\big(Ak(\tau)-|\rho|\big)^2 + \big(1-|\rho|^2\big),
\end{equation}
where $A=\tfrac{1}{\mathcal{R}}(\tfrac{\sigma_w}{w})$ is the ratio of relative width noise to the zero-delay RIN. In Eq.\,\eqref{enhan}, we have considered the mechanical jitter to be a  negligible quantity ($\sigma_\tau\sim0$). This is reasonable as we are using a high-precision motorized translation stage. The enhancement factor $\varepsilon(\tau)$ is used as the primary observable because it is insensitive to any multiplicative scaling of the signal level. This makes it robust against shot-to-shot variations in optical alignment, detector gain, BBO phase-matching efficiency, and the delay-dependent autocorrelation envelope. By normalizing the measured variance to the local mean ($\langle S(\tau)\rangle$), $\varepsilon(\tau)$ cleanly separates the breathing contribution from the underlying amplitude noise baseline, independently of the absolute signal level at each delay.}

\textcolor{black}{For a soliton governed by the area theorem, $\varepsilon(\tau)$ is suppressed below unity at intermediate delays. This produces a characteristic W-shaped dip, which can be identified as a direct signature of soliton dynamics. Also, by analyzing the depth and location of the dip, we can isolate and quantify both the correlation coefficient $\rho$ and the pulse-width breathing $\sigma_w$ independently of time-averaged diagnostics.}

\section{Experimental setup and Methods}\label{sec3}
\begin{figure}[h]
    \centering
    \begin{minipage}{0.48\textwidth}
        \centering
        \includegraphics[width=\linewidth]{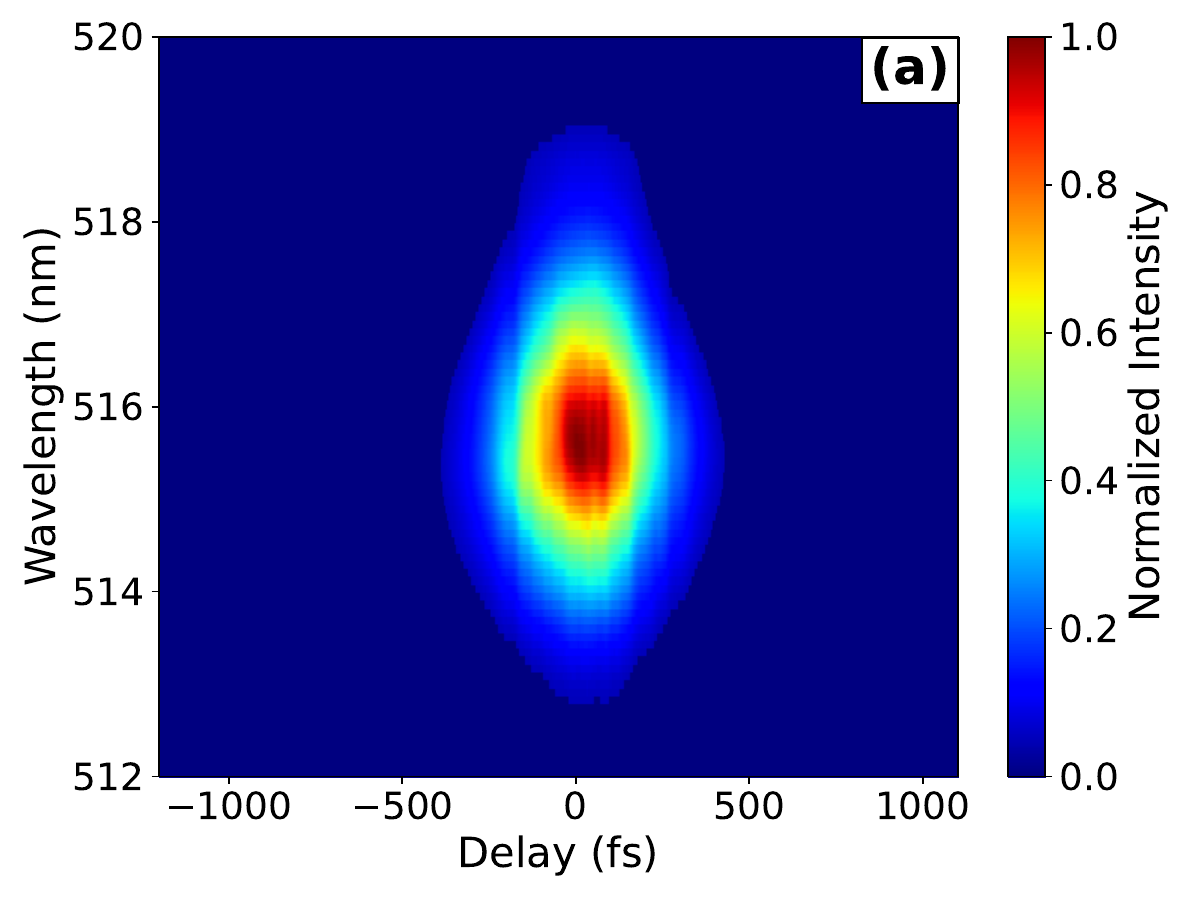}
    \end{minipage}
    \begin{minipage}{0.48\textwidth}
        \centering
        \includegraphics[width=\linewidth]{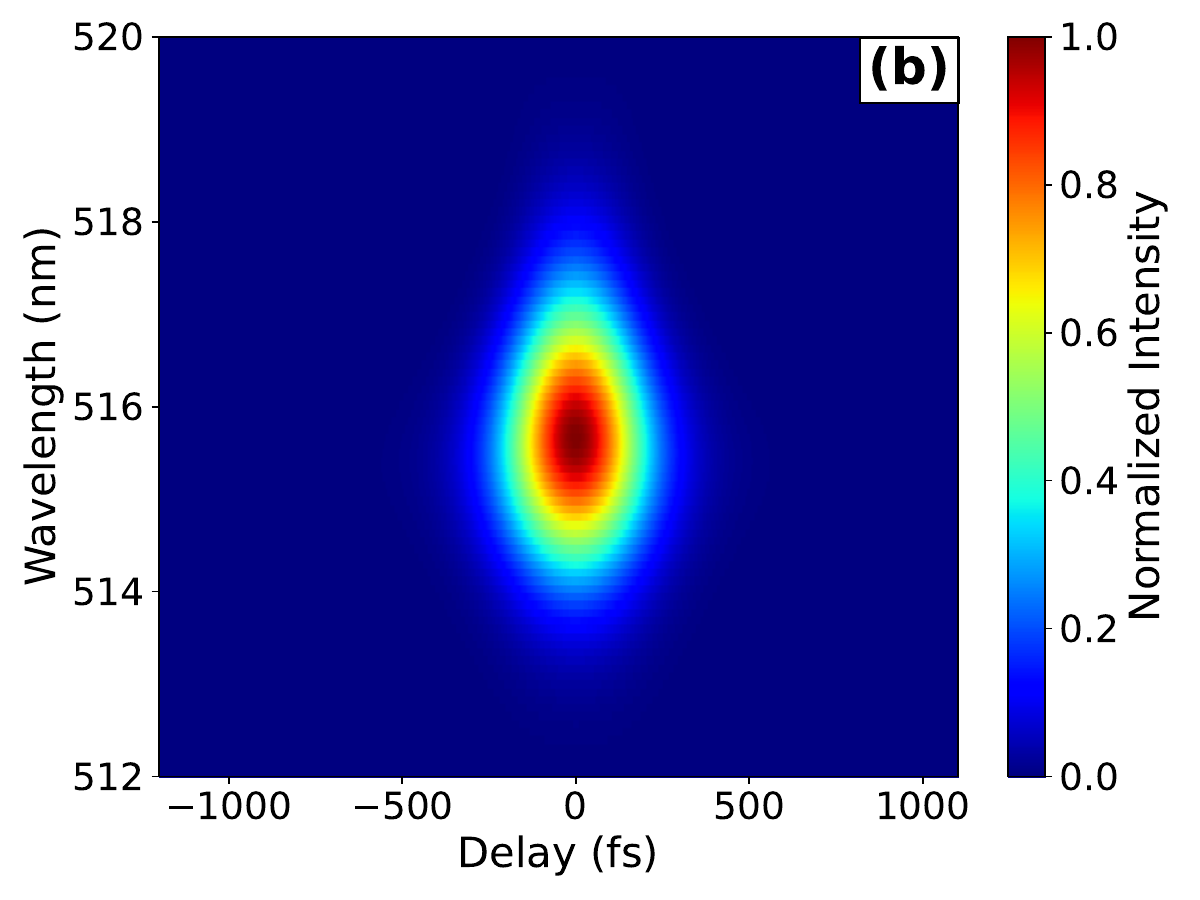}
    \end{minipage}
    \begin{minipage}{0.48\textwidth}
        \centering
        \includegraphics[width=1.0\linewidth]{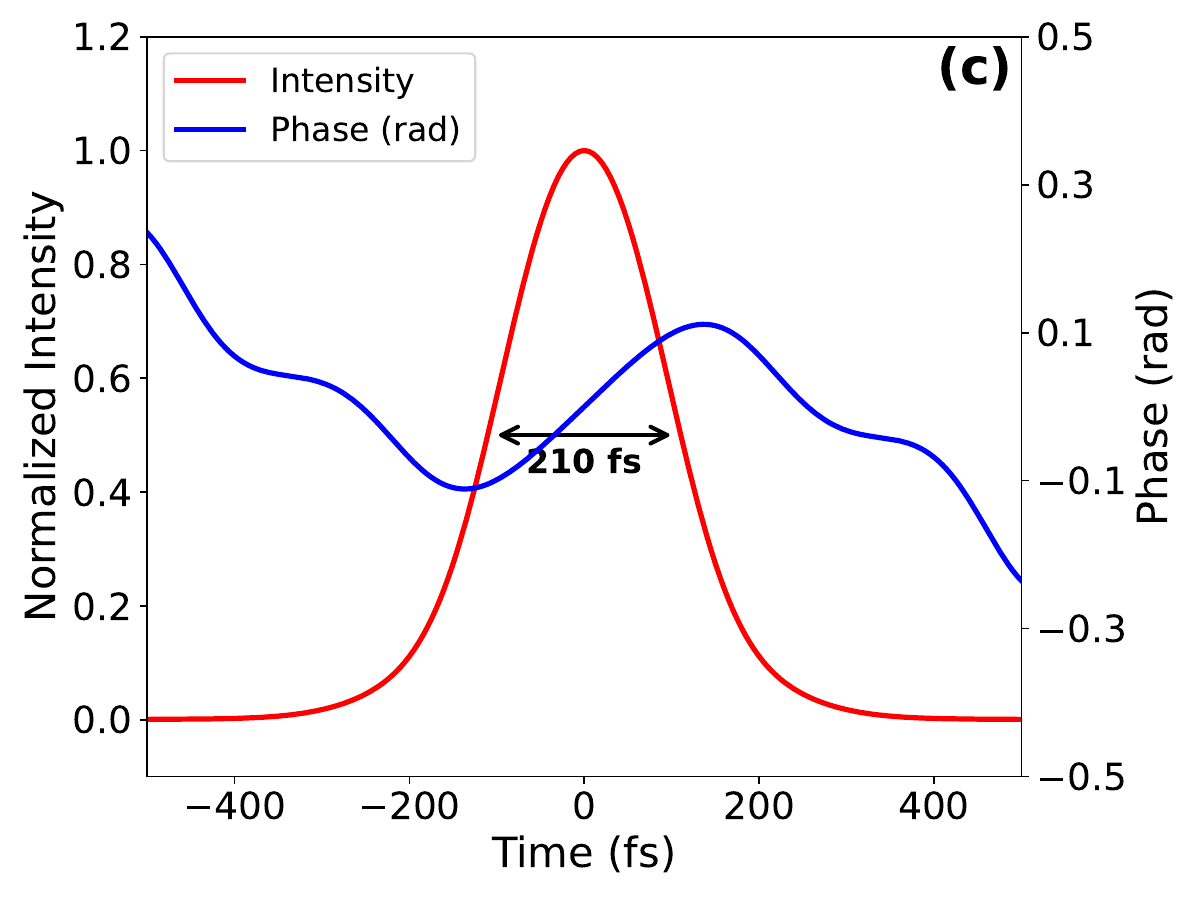}
    \end{minipage}
    \hspace{-0.5cm}
        \begin{minipage}{0.48\textwidth}
        \centering
        \includegraphics[width=0.9\linewidth,height=0.73\linewidth]{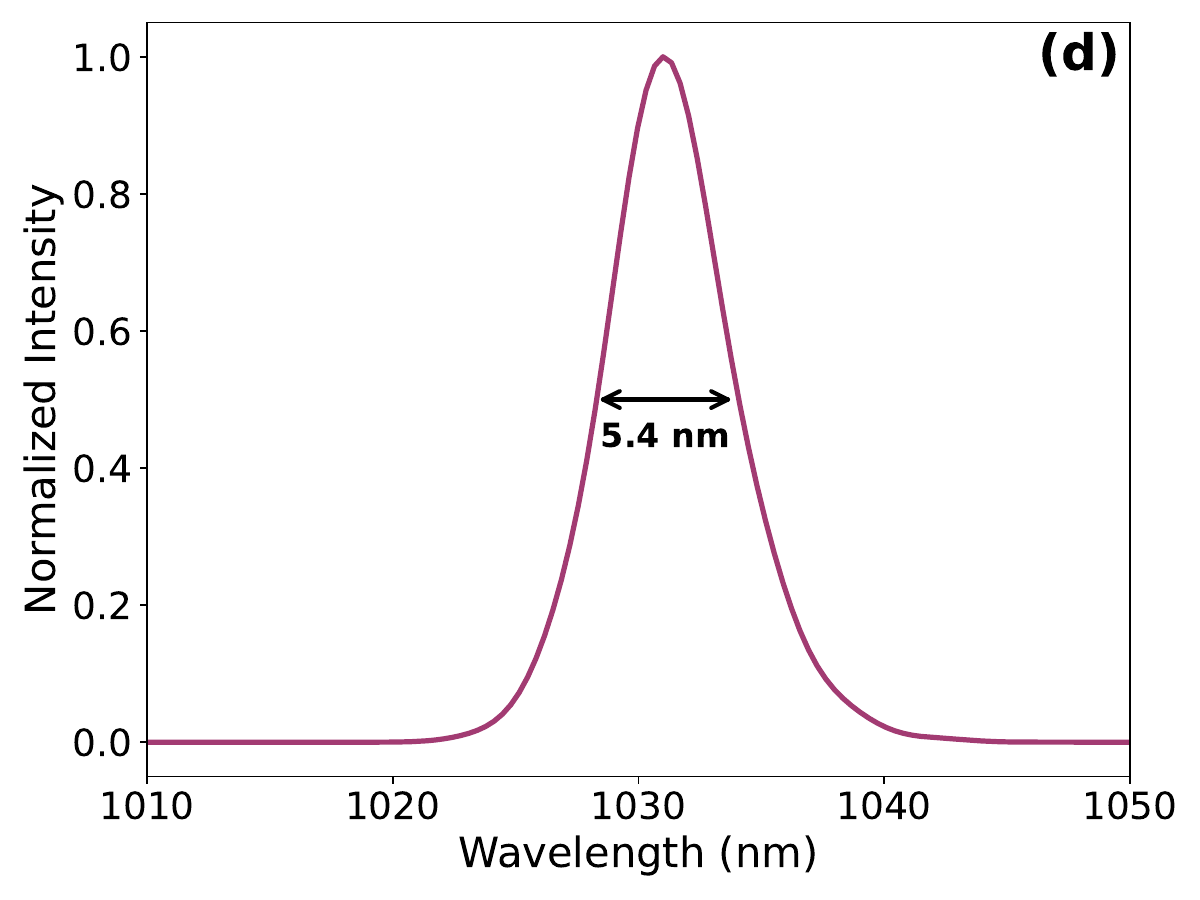}
    \end{minipage}
    \caption{(a) Experimental second harmonic generation frequency resolved optical gating (SHG-FROG) measured directly from the laser output (MENHIR-1030). (b) Reconstructed SHG-FROG trace calculated from the reconstructed pulse field. The low retrieval error (G-error$<0.5\%$) indicates excellent convergence between the measured and calculated spectrograms. (c) Temporal profile of the reconstructed pulse along with the retrieved phase. (d) Optical spectrum retrieved from the SHG-FROG trace by the COPRA algorithm\,\cite{Geib:19} (central wavelength $\sim$1030\,nm, FWHM $\sim$5.4\,nm).}\label{fig2}
\end{figure}
The experiments were performed using commercial mode-locked Yb-doped oscillators. The first laser is housed in a hermetically sealed enclosure and provides free-space output at a central wavelength of 1030 nm with a repetition rate of 160 MHz. It delivers an average power of 200 mW, corresponding to a pulse energy of $\sim$1.25 nJ, with a nominal pulse duration of $\sim200$\,fs  (Menhir Photonics, MENHIR-1030). \textcolor{black}{The second laser is also housed in a hermetically sealed enclosure having a central wavelength of 1045 nm with a repetition rate of 63 MHz. It delivers an average output power of 1.5 W with a pulse duration of $\sim 200\,$fs (Spectra-Physics HighQ-2).}

Initial pulse characterization was conducted using a home-built non-collinear Second Harmonic Generation Frequency-Resolved Optical Gating (SHG-FROG) apparatus (Fig.\,\ref{fig1})\,\cite{DeLong:94}. The output beam is divided by a 50:50 non-polarizing beamsplitter. One arm traverses a free space delay line equipped with a piezo motor linear stage to introduce a variable time delay $\tau$, while the other beam path remains fixed. The two beams are then focused non-collinearly into a BBO crystal, generating a background-free sum-frequency generation (SFG) signal. This signal is spectrally resolved using a fiber-coupled spectrometer (Ocean Optics USB4000).
Unlike a simple intensity autocorrelation, the SHG-FROG technique spectrally resolves the autocorrelation signal at each delay step, producing a two-dimensional spectrogram (Fig.\,\ref{fig2}\,(a)) that encodes both the amplitude and phase of the pulse. The pulse intensity and phase were jointly retrieved from this spectrogram using the common pulse retrieval algorithm (COPRA)\,\cite{Geib:19}, an iterative algorithm that reconstructs the complex electric field by minimizing the difference between the measured and mathematically generated spectrograms (Fig.\,\ref{fig2}\,(b)).
The retrieval converged to a FROG error (G-error) of $<0.5\%$, confirming the fidelity of the reconstructed pulse shape for both the lasers. For the 1030 nm laser, the retrieved temporal profile fits a clean sech$^2$ shape with a full-width at half-maximum (FWHM) of 210\,fs. The time-bandwidth product of $\sim0.32$ is within $\sim2\%$ of the transform limit (0.315) of sech$^2$ pulses, and the retrieved spectral phase is flat to within $\pm0.1$rad across the pulse bandwidth (Fig.\,\ref{fig2}). These results confirm soliton-like operation with balanced intracavity dispersion, validating the sech$^2$ assumption used in our variance model and establishing the mean pulse width $\tau_0=210$\,fs for subsequent analysis.
\textcolor{black}{Similarly, FROG retrieval of the second laser also confirmed a sech$^2$ temporal profile with a pulse duration of $\sim200$\,fs (FWHM).}

To quantify pulse breathing, we implemented a high-dynamic-range statistical autocorrelation technique using the same non-collinear SHG apparatus described above (Fig.\,\ref{fig1}), but with the fiber-coupled spectrometer replaced by a high-speed avalanche photodiode (Menlo Systems APD210) to enable single-pulse detection at the full 160\,MHz repetition rate. \textcolor{black}{A highly stable motorized translation stage ((CONEX-AG-LS25-27P) was utilized to scan the optical delay $\tau$.} At each delay step, the SHG signal was digitized by a Tektronix 5 Series MSO using FastFrame segmented memory mode, capturing a statistical ensemble of 62,500 pulses. The acquisition was performed using a real-time sampling rate of 6.25\,GS/s. \textcolor{black}{We then computed the numerical integral of the electrical pulse (background-subtracted) for each shot. Because the signal is hardware-bandlimited to 1\,GHz and oversampled (Nyquist rate = 2\,Gs/s), this discrete summation (Eq.\,\eqref{dissum}) provides a jitter-free, robust measure of the total charge generated by the photodiode, which is proportional to the single-shot SHG pulse energy.}


\textcolor{black}{Careful calibration was performed to isolate the laser dynamics from detection noise. The APD signal level was maintained at $\sim$50\% of the saturation at zero delay to maximize the signal-to-noise ratio while ensuring operation strictly within the detector's linear regime. Detector dark noise was characterized with the beam blocked. This dark baseline variance was directly subtracted from the raw measured variance at each delay step to isolate purely optical fluctuations.}
Finally, the delay-dependent statistics (mean $\langle S(\tau)\rangle$ , variance $\langle \Delta S(\tau)^2 \rangle$, and enhancement factor $\varepsilon(\tau)$) were computed for the full autocorrelation trace.

\section{Results}
\begin{figure}[h]
    \centering
    \begin{minipage}{0.48\textwidth}
        \centering
        \includegraphics[width=\linewidth]{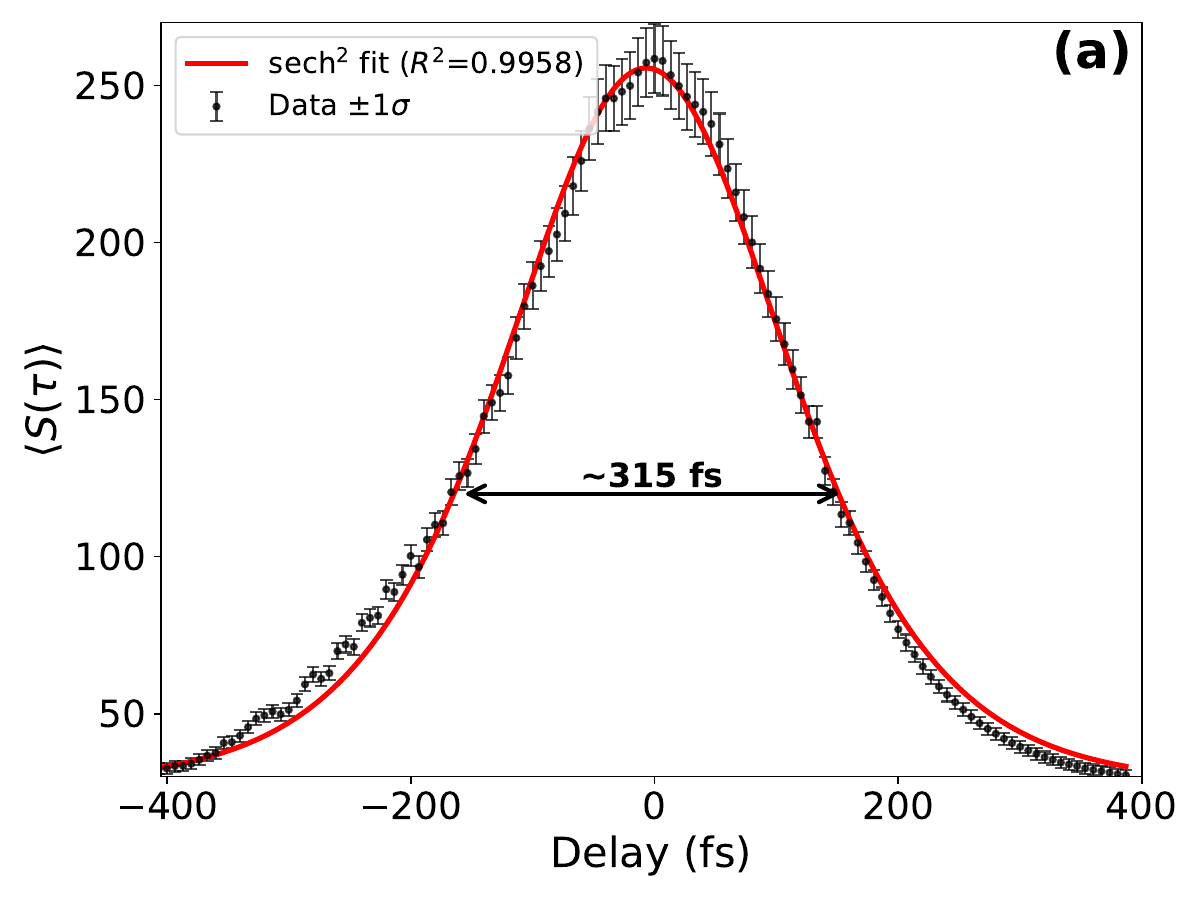}
    \end{minipage}
    \begin{minipage}{0.48\textwidth}
        \centering
        \includegraphics[width=\linewidth]{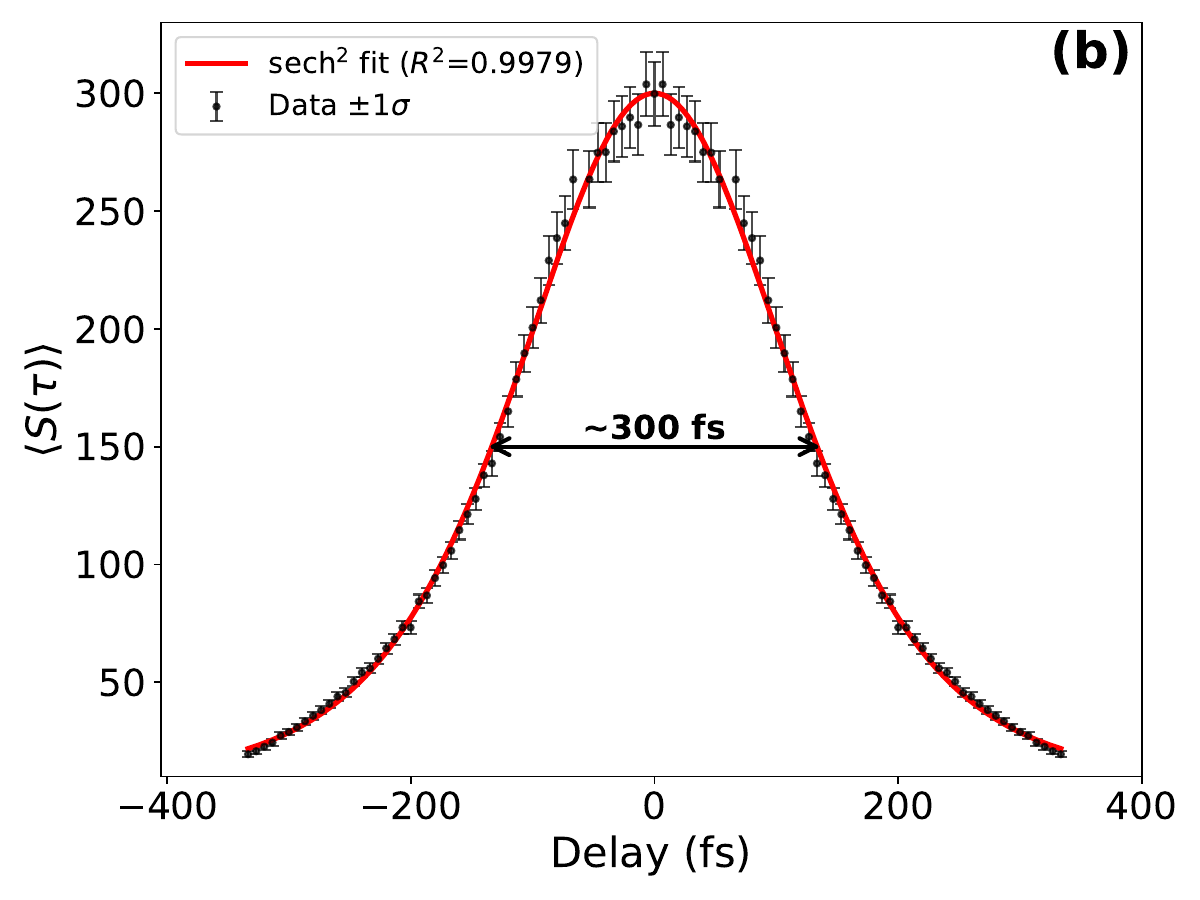}
    \end{minipage}
    \begin{minipage}{0.48\textwidth}
        \centering
        \includegraphics[width=0.98\linewidth,height=5cm]{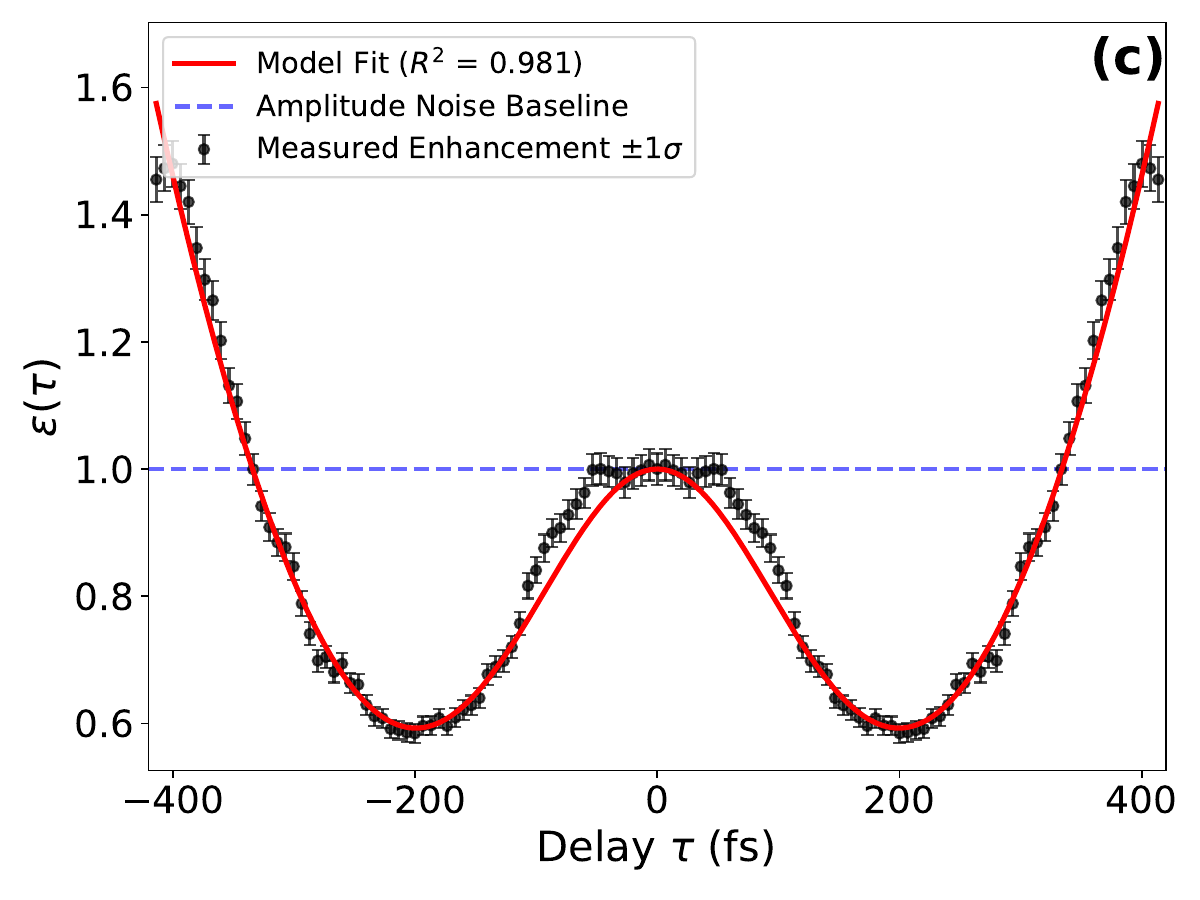}
    \end{minipage}
        \begin{minipage}{0.48\textwidth}
        \centering
        \includegraphics[width=\linewidth]{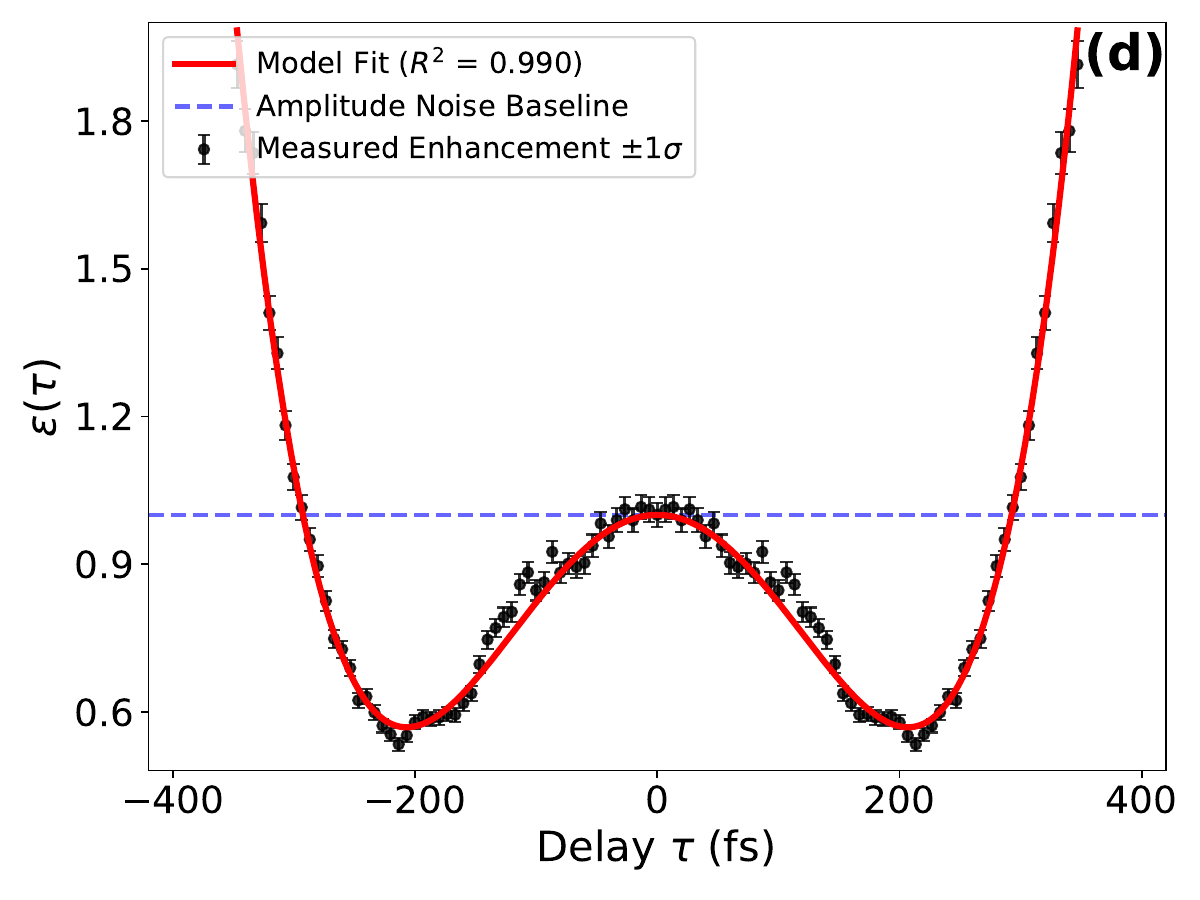}
    \end{minipage}
    \caption{\textcolor{black}{Extraction of pulse breathing dynamics: (a, b) Mean integrated SFG autocorrelation signal ($S(\tau)$) versus delay for the 1030 nm and 1045 nm lasers, respectively. The solid red line represent the corresponding sech$^2$ fits ($R^2\sim0.995$), yielding autocorrelation FWHMs of $\sim315$\,fs and $\sim300$\,fs. (c, d) Corresponding Fano enhancement factor $\varepsilon(\tau)$ for both cases. The experimental data exhibit a characteristic ``W-shaped'' profile, which is a signature of both pulse breathing and anti-correlated amplitude and width fluctuations ($\rho<0$), consistent with the soliton area theorem. The solid red curve represents the fit using Eq.\,\eqref{enhan}, enabling the simultaneous extraction of $\rho$ and the intrinsic pulse-width breathing parameter ($\sigma_w/w$).}}\label{fig3}
\end{figure}
We first analyze the integrated autocorrelation profile as a function of delay, $(\langle S(\tau)\rangle)$, shown in Fig.\,\ref{fig3}(a) and \ref{fig3}(b) for the two lasers. The data fits well to a sech$^2$ profile ($R^2>0.995$) with a FWHM$\sim315\,$fs and $\sim 300\,$fs. These values are consistent with the pulse duration retrieved via SHG-FROG. 
\textcolor{black}{At zero delay, the sensitivity terms in Eq.\,\eqref{eq:variance_prop} vanish, and the variance reduces to pure amplitude noise, $\langle\Delta S(0)^2\rangle = \mathcal{R}^2\langle S(0)\rangle^2$. Over the statistical ensemble of $N=62,500$ pulses, the relative intensity noise (RIN) baseline is calibrated to $\mathcal{R}=4.20(2)\%$ for the 1030 nm laser and $\mathcal{R}=4.50(2)\%$ for the 1045 nm laser. The small statistical uncertainty ensures that the baseline amplitude noise floor is tightly constrained for the subsequent extraction of pulse breathing dynamics.}

\textcolor{black}{To isolate the pulse breathing, we compute the Fano enhancement factor $\varepsilon(\tau)$ across the delay axis (Fig.\,\ref{fig3}(c) and \ref{fig3}(d)). In the presence of uncorrelated or positively correlated amplitude and width fluctuations ($\rho\geq0$), the enhancement factor reduces to $\varepsilon(\tau)\geq1+(Ak(\tau))^2$. This produces a broad ``M-shaped'' excess variance peak at the autocorrelation shoulders. Instead, our experimental data shows a ``W-shaped'' profile, dropping to $\sim0.6$ of the baseline amplitude noise floor. This is a signature of destructive interference between the amplitude and width noise channels, as can be seen from the negative cross-term in Eq.\,\eqref{eq:variance_prop}. It explicitly proves that the lasers are operating in a coupled soliton regime, where any increase in pulse energy is simultaneously met with a narrowing of the pulse width\,\cite{AGRAWAL2013129}.}

\textcolor{black}{To quantitatively extract this dynamics, we fit the measured enhancement profile to the analytical model in Eq.\,\eqref{enhan}. For the 1030 nm laser, the fit extracts a relative breathing magnitude of $\sigma_w/w=1.56(6)\%$. For a 210 fs pulse, this corresponds to a width fluctuation of 3.2(1) fs. Also, the depth of the enhancement dip directly constrains the amplitude-width correlation coefficient, yielding $\rho=-0.70(8)$. Similarly, for the 1045 nm laser, the fit gave $\sigma_w/w=1.43(1)\%$, corresponding to a width fluctuation of 2.84(2)\,fs. For this laser, the correlation coefficient turned out to be $\rho=-0.66(1)$.}

\textcolor{black}{To quantify the fit quality for the enhancement data, we computed the root-mean-square error (RMSE). The low RMSE ($\sim$0.03), comparable to the mean per-point measurement uncertainty, confirms that the residuals are consistent with shot noise, with no systematic structure exceeding the measurement uncertainty. Empirical validation of the $\sigma_\tau\sim0$ approximation is provided by the agreement (within uncertainty) between the measured enhancement factor $\varepsilon(\tau)$ and the zero-stage-noise model prediction across the full delay range.}

\section{Discussion}
\textcolor{black}{The central result of this work is the observation of a characteristic W-shaped enhancement Fano profile in both oscillators. As established by Eq.\,\eqref{enhan}, the dip below unity cannot arise from amplitude noise, uncorrelated breathing, or stage noise. Hence the W-shape is a model-independent signature of anti-correlated  soliton dynamics. The extracted correlation coefficients, $\rho=-0.70(8)$ for the 1030 nm laser and $\rho=-0.66(1)$ for the 1045 nm laser, confirm this anti-correlation. The deviation from the ideal limit ($\rho=-1$) represents the presence of competing noise sources. One such source is the contribution from the spontaneous emission (ASE) from the gain medium, which adds random energy perturbations that are uncorrelated with the pulse width changes.}

\textcolor{black}{The extracted pulse-width breathing, $\sigma_w=3.2(1)\,$fs (1030 nm) and $\sigma_w=2.86(2)\,$fs (1045 nm), represents the intrinsic pulse stability of each oscillator. These are inaccessible to time-averaged diagnostics.}
This pulse breathing has \textcolor{black}{qualitative} implications for nonlinear applications, particularly supercontinuum generation in photonic crystal fibers (PCF).
\textcolor{black}{While both oscillators show consistent anti-correlated soliton dynamics, we focus the following application analysis on the 1030 nm laser, as it forms the seed source for our ongoing development of an optical frequency comb for precision spectroscopy of metastable helium\,\cite{PhysRevLett.125.013002,doi:10.1126/science.abk2502,doi:10.1126/science.adj2462,doi:10.1126/science.adj2610,PhysRevLett.134.223001}.}
For a pump wavelength of 1030\,nm launched into an anomalous dispersion PCF, the spectral broadening is governed by soliton fission\,\cite{PhysRevLett.88.173901,RevModPhys.78.1135}, a process critically dependent on the input soliton order $N$\,\cite{RevModPhys.78.1135}:
\begin{equation}\label{soliton_order}
    N=\sqrt{\frac{\gamma P_0\tau_0^2}{\vert\beta_2\vert}},
\end{equation}
where $\gamma$ is the nonlinear parameter of the fiber, $P_0$ is the peak power of the laser, $\tau_0$ is the pulse width and $\beta_2$ is the group velocity dispersion parameter. Assuming the pulse energy is clamped by gain saturation ($E_p \approx \text{constant}$), the peak power scales inversely with width ($P_0 \propto 1/\tau_0$). Substituting this into Eq.\,\eqref{soliton_order} reveals that the soliton order scales with the square root of the pulse duration: $N \propto \sqrt{\tau_0}$. Following established scaling laws for the soliton fission regime\,\cite{AGRAWAL2013129}, the bandwidth of the generated supercontinuum increases linearly with $N$. \textcolor{black}{Consequently, our measured breathing in pulse width provides an indication that supercontinuum bandwidth fluctuations may occur at the percent level, although a more quantitatively accurate prediction would require detailed Generalized Nonlinear Schr\"{o}dinger Equation (GNLSE) simulations seeded by the full measured pulse statistics\,\cite{RevModPhys.78.1135,corwin2003fundamental}.} While this percentage appears small, these fluctuations manifest as amplified relative intensity noise at the spectral edges of the continuum, where the intensity slope is steepest. This flickering of the bandwidth will degrade coherence and limits the signal-to-noise ratio in spectroscopic applications\,\cite{PhysRevLett.90.113904,corwin2003fundamental}. Unlike timing jitter, which can be actively stabilized, these breathing dynamics are intrinsic to the oscillator’s soliton shaping and often pass undetected by standard diagnostics. This suggests that for demanding applications, prioritizing low-breathing oscillators (which can be achieved by using a pump source with extremely low RIN or actively stabilizing the pump power before it enters the laser cavity and also by optimizing the cavity's noise transfer function) is as critical as minimizing average intensity noise.

To place this in perspective, we compare the measured width fluctuations ($\sigma_w \approx 3.3$\,fs) to the laser's timing jitter. Timing jitter, typically characterized by integrated phase noise, measures arrival time stability but provides no information about pulse shape. For context, the manufacturer specifies $\sim$7 fs of integrated timing jitter over a 1 kHz to 1 MHz analysis bandwidth (Menhir Photonics datasheet), capturing the dominant acoustic and mechanical noise sources. This corresponds to only $\sim 3\%$ of the pulse duration. The breathing width fluctuations above are smaller than the magnitude of the timing fluctuations. The unique value of statistical autocorrelation lies in identifying this $\sigma_w$ component, offering a direct probe into the cavity noise transfer function that timing jitter measurements miss.

\begin{figure}[ht]
    \centering
    \begin{minipage}{0.48\textwidth}
        \centering
        \includegraphics[width=\linewidth]{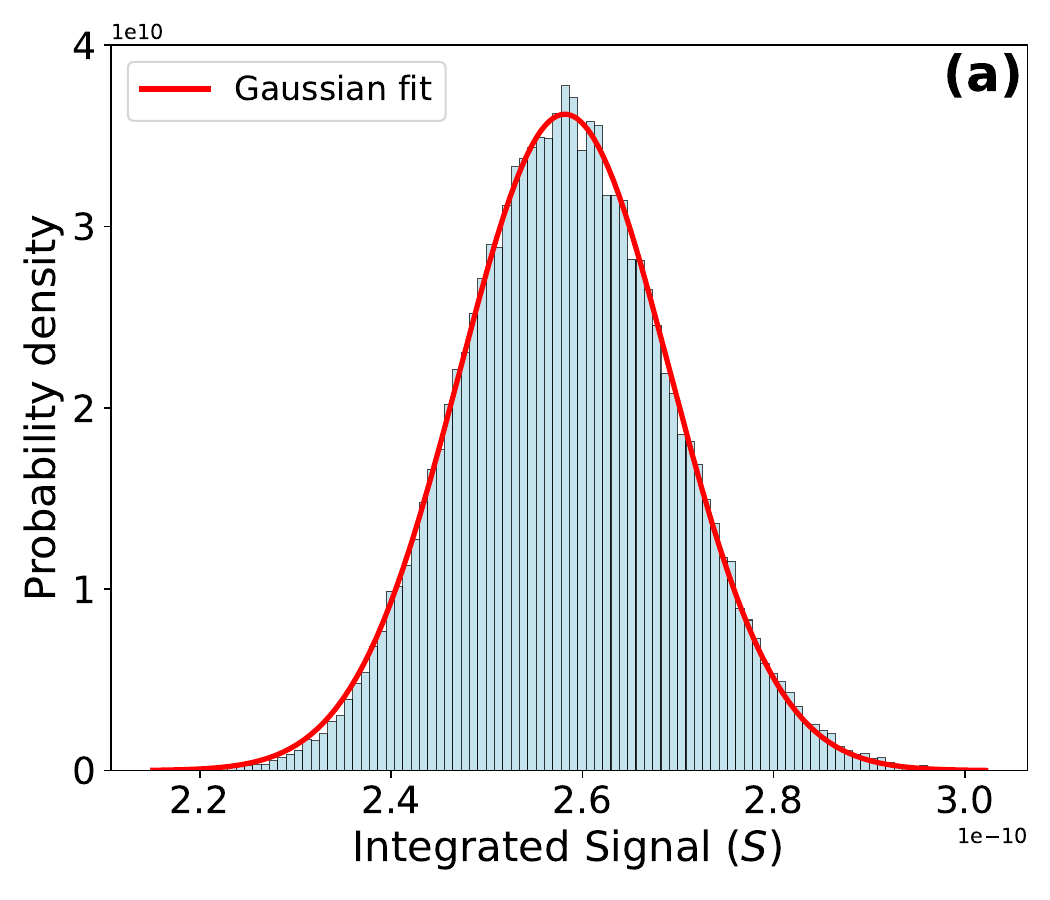}
    \end{minipage}
    \begin{minipage}{0.48\textwidth}
        \centering
        \includegraphics[width=\linewidth]{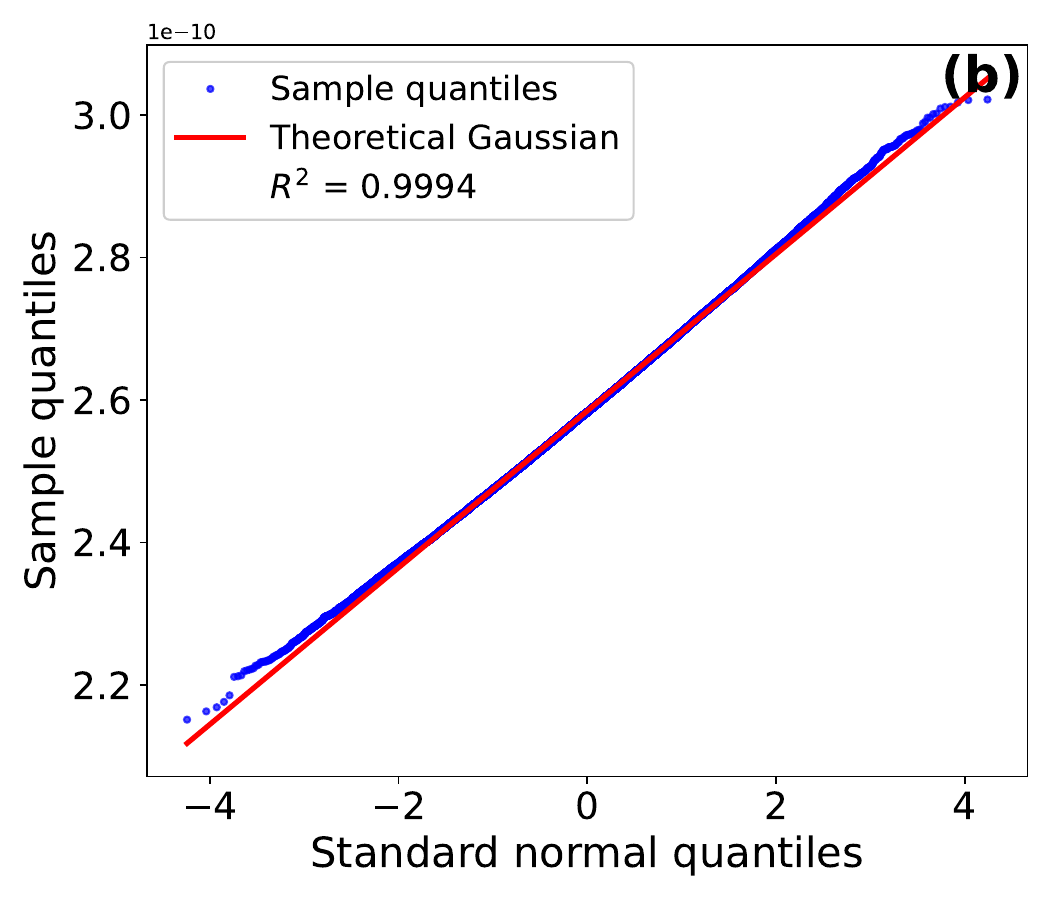}
    \end{minipage}
    \centering
    \begin{minipage}{0.48\textwidth}
        \centering
        \includegraphics[width=\linewidth]{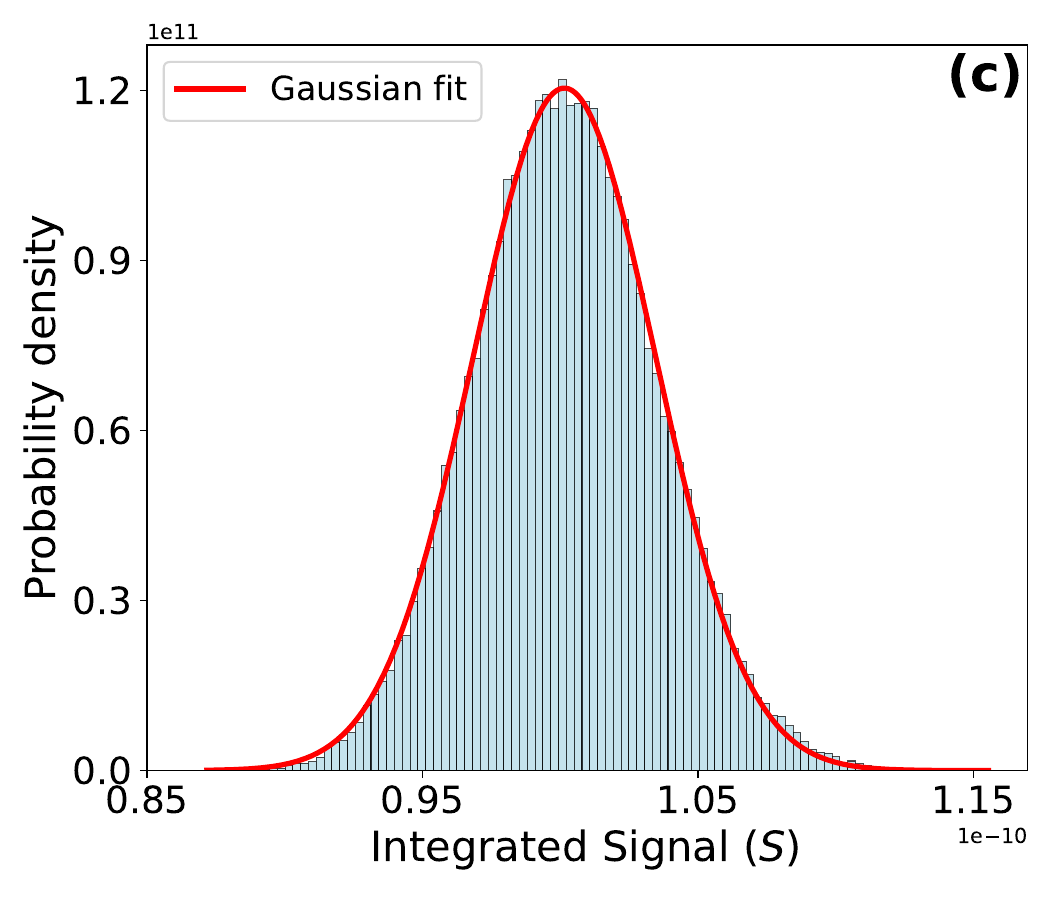}
    \end{minipage}
        \begin{minipage}{0.48\textwidth}
        \centering
        \includegraphics[width=\linewidth]{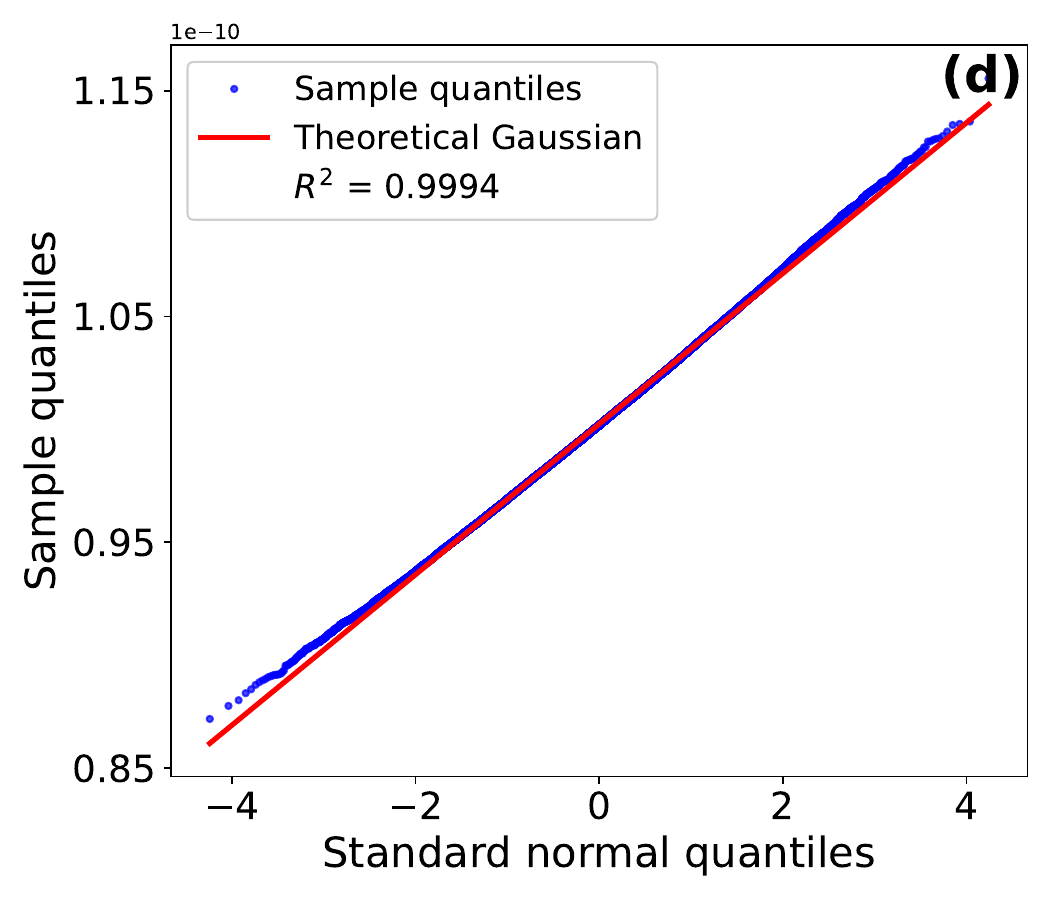}
    \end{minipage}
    \caption{ Validation of Gaussian statistics assumption (1030 nm laser). Voltage distributions at two representative delays: (a-b) $\tau=0$\,fs and (c-d) $\tau=200\,$fs (a) and (c): Histograms with Gaussian fits (red). (b) and (d): Quantile-Quantile plots showing excellent linearity ($R^2>0.996$), confirming normality.}\label{fig5}
\end{figure}
To validate the Gaussian statistics assumption implicit in our variance decomposition (Eq.\,\eqref{eq:variance_prop}), we examined the integrated voltage distributions at multiple delay points. Fig.\,\ref{fig5} shows histograms and Q-Q plots for the 1030 nm laser at $\tau=0$ (no delay) and 200\,fs (near the minimum of $\varepsilon(\tau)$). All distributions exhibit excellent agreement with Gaussian fits, with Q-Q plots (sample quantiles against theoretical quantiles from a standard normal distribution) showing strong linearity $(R^2>0.99)$. This validates the use of the second-order variance decomposition formula derived from first-order error propagation. This is physically motivated by the Central Limit Theorem, since laser fluctuations arise from many independent noise sources. This can be verified experimentally by examining the voltage distribution at each delay.
For lasers exhibiting non-Gaussian dynamics-such as Q-switched mode-locking or rogue wave generation\,\cite{solli2007optical,Navitskaya:22,teugin2023real}, higher-order cumulants beyond the variance would be needed to fully characterize the pulse statistics.
The technique is currently limited to pulse widths $\tau$ much longer than the SHG response time (effectively instantaneous for femtosecond pulses) and delays where SNR remains adequate ($\vert\tau\vert < 5\tau$ typically).
The variance decomposition assumes a sech$^2$ autocorrelation profile, validated by the FROG retrieval (R$^2 = 0.995$, TBP within 2\% of the transform limit). Static deviations from sech$^2$ do not generate excess variance; only shot-to-shot fluctuations contribute.
\textcolor{black}{Because the SHG autocorrelation signal depends only on the temporal intensity profile $I(t)=|E(t)|^2$, the technique is fundamentally insensitive to purely phase-based linear chirp fluctuations. Furthermore, the SHG-FROG retrieval helped in ruling out any higher-order dynamic chirp.}
Future work could extend the model to include chirp fluctuations by adding a $(\partial S/\partial C)^2\sigma_C^2$ term, where $C$ is the chirp parameter, or apply it to fiber lasers where competing noise mechanisms (ASE, Raman scattering) may alter the correlation structure.
\textcolor{black}{For lasers with significant stable asymmetry or non-sech$^2$ shapes, the variance decomposition framework of Eq.\,\eqref{eq:variance_prop} remains exactly valid. Because standard mode-locked pulses generally maintain a self-similar functional form, the sensitivity factor $k(\tau)$ can be computed purely numerically from the slope of the measured mean autocorrelation trace, utilizing the self-similarity property $\partial S_{ac}/\partial w =-(\tau/w)\partial S_{ac}/\partial \tau$, without assuming any specific pulse shape. For lasers where pulse asymmetry fluctuates independently shot-to-shot, this would constitute an additional noise channel. However, for our stable oscillators, the FROG retrieval confirms a consistent pulse shape, making such fluctuations negligible.}

\textcolor{black}{It is instructive to compare this method with the Dispersive Fourier Transform (DFT)\,\cite{goda2013dispersive}, which is the standard technique for single-shot spectral dynamics. While DFT is highly effective for observing large scale instabilities (e.g. rogue waves), it faces distinct challenges for measuring breathing in stable, nanojoule oscillators at 1030 nm. DFT relies on kilometers of dispersive fiber, which at 1030 nm introduces significant propagation losses, often requiring optical pre-amplification that can mask intrinsic cavity noise. Furthermore, extracting temporal breathing from DFT spectra requires assumptions about the instantaneous phase\,\cite{goda2013dispersive}, whereas our statistical autocorrelation method operates directly in the time domain. By avoiding external amplification and dispersive propagation, our statistical autocorrelation method achieves the high dynamic range required to resolve temporal width fluctuations using only a standard SHG autocorrelation apparatus, making it wavelength independent and immediately applicable wherever a mode-locked oscillator and a suitable nonlinear crystal are available.}

\section{Conclusion}
We have demonstrated that statistical analysis of SHG autocorrelation reveals pulse breathing dynamics in mode-locked lasers. Applied to our\,1030 nm laser, we measured pulse width fluctuations of $\sim3.2$\,fs\,($\sim1.6\%$). The method is inexpensive, fast, requires only standard apparatus, and provides information inaccessible to conventional FROG or spectral measurements. The measured breathing amplitude enables a qualitative prediction of the impact on supercontinuum spectral stability. This technique fills a gap between time-averaged pulse characterization and single-shot spectroscopy, offering practical access to cavity noise dynamics for oscillator development and application optimization.

This diagnostic capability will be important in our ongoing development of an optical frequency comb at 1030 nm. This will be used for precision spectroscopy of atomic transitions, using our ultracold metastable $^3$He and $^4$He\,\cite{PhysRevA.107.033313} to improve precision spectroscopic measurements\,\cite{PhysRevLett.125.013002,doi:10.1126/science.abk2502,doi:10.1126/science.adj2462,doi:10.1126/science.adj2610,PhysRevLett.134.223001}. 

\textcolor{black}{Although the present implementation assumes a sech$^2$ pulse shape appropriate for soliton oscillators, the underlying variance decomposition relies on the self-similarity property $\frac{\partial I}{\partial w} = -\frac{\tau}{w}\frac{\partial I}{\partial \tau}$. Consequently, the framework is immediately applicable to any stable pulse shape (shot-to-shot stability), such as the Gaussian profiles characteristic of dispersion-managed lasers\,\cite{Haus:93}, where the characteristic variance peaks simply map to the new spatial geometry. For dissipative solitons or pulses with complex shapes, the sensitivity factors $k(\tau)$ and $j(\tau)$ can be computed numerically from the measured mean autocorrelation profile rather than analytically, preserving the diagnostic utility of the W-shaped enhanced Fano profile.}
For pulses with unknown or complex shapes, the method remains useful as a qualitative indicator of breathing dynamics through the enhancement factor $\varepsilon$, even where quantitative extraction of $\sigma_{w}$ is not possible.
\begin{backmatter}

\bmsection{Funding}
This work was supported through Australian Research
Council Discovery Project Grant Nos. DP240101346 and DP240101441. S.S.H.
was supported by Australian Research Council Future Fellowship
Grant No. FT220100670. S. K. was supported
by an Australian Government Research Training Program
scholarship.
\bmsection{Acknowledgment}
The authors thank Prof.\,Barry Luther-Davies for fruitful discussions and for lending the 1045 nm laser.
\bmsection{Disclosures}
The authors declare no conflicts of interest.

\bmsection{Data availability} Data and code underlying the results presented in this paper are available at \cite{pulse_breth_data}.

\end{backmatter}

\bibliography{sample}

\end{document}